\documentclass[aps,prd,onecolumn,floats,floatfix,letterpaper,nofootinbib,preprintnumbers,superscriptaddress]{revtex4}

\usepackage{amssymb,amsmath,latexsym,mathrsfs}
\usepackage{graphicx}
\usepackage{epsfig}
\usepackage{multirow}
\usepackage{array}
\usepackage{xspace}
\usepackage{color,xcolor}
\usepackage{isotope}
\usepackage[normalem]{ulem}
\usepackage[colorlinks=true
,urlcolor=blue
,anchorcolor=blue
,citecolor=blue
,filecolor=blue
,linkcolor=blue
,menucolor=blue
,linktocpage=true
,pdfproducer=medialab
,pdfa=true
]{hyperref}

\usepackage{epsfig,psfrag,rotating,soul}
\usepackage{rotfloat}
\usepackage{slashed}
\usepackage[normalem]{ulem}

\allowdisplaybreaks

\begin{document}

\title{Search for Light Exotic Fermions in Double-Beta Decays}

\author{Matteo Agostini}
\email{matteo.agostini@ucl.ac.uk}
\affiliation{Department of Physics and Astronomy, University College London, Gower Street, London WC1E 6BT, UK. }
\affiliation{Physik-Department, Technische Universit\"at M\"unchen, James-Franck-Stra\ss e, 85748 Garching, Germany}
\author{Elisabetta Bossio}
\email{elisabetta.bossio@tum.de}
\author{Alejandro Ibarra}
\email{ibarra@tum.de}
\author{Xabier Marcano}
\email{xabier.marcano@tum.de}
\affiliation{Physik-Department, Technische Universit\"at M\"unchen, James-Franck-Stra\ss e, 85748 Garching, Germany}

\begin{abstract}
The Standard Model of Particle Physics predicts the double-$\beta$ decay of certain nuclei with the emission of two active neutrinos. In this letter, we argue that double-$\beta$ decay experiments could be used to probe models with light exotic fermions through the search for spectral distortions in the electron spectrum with respect to the Standard Model expectations. We consider two concrete examples: models with light sterile neutrinos, singly produced in the double-$\beta$ decay, and models with a light $Z_2$-odd fermion, pair produced due to a $Z_2$ symmetry. We estimate the discovery potential of a selection of double-$\beta$ decay experiments and find that future searches will test for the first time a new part of the parameter space of interest at the MeV-mass scale.

\end{abstract}

\preprint{TUM-HEP 1306/20}

\newcommand{\nnbb}   {\ensuremath{2\nu\beta\beta}}
\newcommand{\oobb}   {\ensuremath{0\nu\beta\beta}}
\newcommand{\nNbb}   {\ensuremath{\nu N\beta\beta}}
\newcommand{\NNbb}   {\ensuremath{NN\beta\beta}}
\newcommand{\XXbb}   {\ensuremath{\chi\chi\beta\beta}}

\maketitle

\section{Introduction}\label{Intro}

Many models of new physics contain new spin 1/2 particles, singlets under the Standard Model gauge group, possibly related to the mechanism of neutrino mass generation, or to the dark matter of the Universe. An archetype of such fermions is the sterile neutrino, also called right-handed neutrino. In a variant of this scenario, the singlet fermion is furnished with a $Z_2$ symmetry so it can  only be produced in pairs. Unfortunately, in either of these scenarios, the mass of the light exotic fermion and the coupling strength to the Standard Model particles are free parameters of the model, leaving a vast parameter space that must be probed in laboratory experiments, or with astrophysical and cosmological observations. 

From the phenomenological standpoint, sub-GeV sterile neutrinos are particularly attractive, since they could be produced in charged lepton or in hadron decays and possibly be discovered in a laboratory experiment. In fact, there is an intense experimental program aiming to detect signals of sterile neutrinos in the sub-GeV range, either in dedicated experiments, or as a by-product of an experiment initially designed for a different purpose. The current limits are fairly stringent below $\sim 100$\,keV and above $\sim 100$\,MeV (for an updated summary of the constraints, see {\it e.g.}~\cite{Bolton:2019pcu}). However, the intermediate mass region remains relatively unexplored and some of the current constraints date back to the 1990s~\cite{Deutsch:1990ut,Schreckenbach:1983cg}. The scenario with $Z_2$-odd singlet fermions has been even less studied, and, to the best of our knowledge, there is no laboratory constraint for the mass range $\sim 100$\,keV $- 100$\,MeV.

In this letter, we explore the possibility of searching for the production of light exotic fermions in double-$\beta$ decays. These decays are nuclear transitions in which the atomic number $Z$ increases by two units while the nucleon number $A$ remains constant. The process is allowed by the Standard Model as long as two neutrinos and two electrons are emitted: $(A,Z) \rightarrow (A,Z+2) + 2e + 2\bar\nu$ (\nnbb\ decay). 
If light exotic fermions exist, they will also be produced in double-$\beta$ decays, replacing  one or both neutrinos in the final state. 
While neutrinos or exotic fermions are challenging to detect, the energy distribution of the electrons emitted in the process can be accurately measured. The shape of this distribution carries information about the full kinematic of the process and can be used to reconstruct which other particles have been emitted.
The fermion mass range accessible with this kind of search extends from a few hundred keV to a few MeV, {\it i.e.}, from the energy threshold of the experiments till the Q-values of the decay.

In the next couple of decades, several experiments with the capability of measuring the electron energy distribution in \nnbb\ decays will get online~\cite{Dolinski:2019nrj}. The primary physics goal of these experiments is the discovery of double-$\beta$ decays with only the two electrons in the final state: $(A,Z) \rightarrow (A,Z+2) + 2e$. Observing this so-called neutrinoless double-$\beta$ (\oobb) decay would have far-reaching implications. It would prove that the lepton number is not a global symmetry and that neutrinos have a Majorana mass component~\cite{Furry:1939qr,Schechter:1981bd}. 
This physics goal already justifies major investments for experimental infrastructure. Nevertheless, there are several other valuable physics searches that can benefit from the  low-background and large target mass of these experiments~\cite{Hirsch:1995in,D_az_2014,Deppisch:2020mxv,Deppisch:2020sqh}.
Our proposed search for light exotic fermions provides a valuable extension of their physics program.

We will discuss and estimate the discovery potential of double-$\beta$ decay experiments under two illustrative scenarios that extend the Standard Model by adding light exotic fermions. In the first one, we assume the existence of a massive sterile neutrino $N$ that mixes with the electron neutrino.
This opens the possibility to observe a double-$\beta$ decay final state in which a standard neutrino is replaced by a sterile neutrino (\nNbb\ decay): $(A,Z) \rightarrow (A,Z+2) + 2e + \nu + N$, with a modified spectrum due to the sterile neutrino mass. 
This is indeed the same principle as the one used in kink searches with single-$\beta$ decays~\cite{Shrock:1980vy,Riis:2010zm,Mertens:2014nha,Abada:2018qok,Aker:2020vrf}, which currently provide the strongest laboratory bounds in our mass range of interest~\cite{Holzschuh:2000nj,Deutsch:1990ut,Schreckenbach:1983cg,Derbin:2018dbu}.
As we will see, the presence of light exotic fermions does not create a kink in the total double-$\beta$ energy spectrum but still creates a continuous distortion that is detectable by \oobb\ decay experiments. 

The second scenario that we will consider is characterized by the presence of a new symmetry that forbids the production of a single exotic fermion in double-$\beta$ decays. This is a typical scenario for models adding new exotic fermions $\chi$ as dark matter candidates, which are charged under a $Z_2$ symmetry to make them stable. 
This kind of models cannot be tested through single-$\beta$ decays, but would result in a new final state in double-$\beta$ decays (\XXbb\ decay). Thus, double-$\beta$ decay experiments can provide the first laboratory-based constraints on several models.

In the following, we will review the phenomenology of the aforementioned scenarios and present our calculations of how the electron energy distribution is affected by the model parameters. The experimental identification of distortions in the electron energy distribution requires a spectral fit whose accuracy is limited by statistical and systematic uncertainties. We will discuss how to set up a statistical analysis and evaluate the impact of systematic uncertainties. Finally, we will derive sensitivity projections for the most promising \oobb\ experiments, assuming the systematic uncertainties and experimental performance achieved by running experiments.

\section{ Double-$\pmb{\beta}$ decay into sterile neutrinos}\label{Models}

One of the simplest extensions of the Standard Model consists in adding to the particle content one spin 1/2 particle, singlet under the gauge group, $N$. The gauge symmetry allows a Yukawa coupling of $N$ to the Standard Model Higgs doublet and lepton doublets, which leads upon electroweak symmetry breaking to a Dirac mass term which couples $N$ to $\nu$, $m_D\overline{\nu} N$. For this reason, $N$ is commonly denominated right-handed neutrino or sterile neutrino. Furthermore, the gauge symmetry also allows a Majorana mass term for $N$, $M \overline{N^c} N$ which we assume to be in the range 0.1-2\,MeV. The two parameters of the model, $m_D$ and $M_N$ are usually recast in terms of the misalignment angle between the interaction and mass eigenstates, $\sin\theta\simeq m_D/M$, and the mass of the heaviest eigenstate $m_N\simeq M$.

If kinematically possible, sterile neutrinos could be produced in any decay process involving Standard Model neutrinos, due to the active-sterile mixing angle. The new decay channels lead to distortions in the energy distribution for the visible particles compared to the Standard Model expectations, and which then constitute a test for the sterile neutrino scenario. In this work we concentrate on the double-$\beta$ decay. In the presence of sterile neutrinos, and provided their mass is below the Q-value of the decay, the decay channels  \nNbb\ and \NNbb\ become kinematically possible.
The differential energy spectrum $\Gamma$ is given by the incoherent superposition of  three channels, and can be expressed as:
\begin{equation}
\frac{d\Gamma}{dT} = \cos^4\theta\, \frac{d\Gamma_{\nu\nu}}{dT}\,  \theta(T_0-T)  + 2\cos^2\theta \sin^2\theta\, \frac{d\Gamma_{\nu N}}{dT} \, \theta(T_0-T-x_N) 
+\sin^4\theta\, \frac{d\Gamma_{NN}}{dT} \, \theta(T_0-T-2x_N)\,, 
\label{eq:dGammadT}
\end{equation}
where $T=(E_{e_1}+E_{e_2}-2m_e)/m_e$ is the sum of the kinetic energies of the two electrons (normalized to the electron mass). This variable is kinematically restricted to be $0\leq T\leq T_0$, $T_0-x_N$ and $T_0-2x_N$, for \nnbb, \nNbb\ and \NNbb\ respectively, with $x_N=m_N/m_e$ and  $T_0=Q_{\beta\beta}/m_e$. $Q_{\beta\beta}$ is the end-point energy assuming massless neutrinos and depends on the particular nucleus.

The \nnbb\ decay has been studied in several works~\cite{Doi:1981mi,Doi:1981mj,Doi:1985dx,Haxton:1985am,Tomoda:1990rs,Kotila:2013gea,Stefanik:2015bya,Simkovic:2018rdz}, always assuming vanishing neutrino masses. In this work we extend this analysis, leaving the masses of the invisible fermions as free parameters. To this end, we follow and generalize \cite{Kotila:2013gea}, and express the differential rate for a generic decay $(A,Z)\rightarrow(A,Z+2)+2e+a+b$, with $a$ and $b$ being either $\nu$ or $N$, as:
\begin{equation}\label{dGammadT}
\frac{d\Gamma_{ab}}{dT} = |\mathcal M_{2\nu}^{\rm eff}|^2\, \frac{dG_{ab}^{(0)}}{dT}\,,\quad 
\end{equation}
with $\mathcal M_{2\nu}^{\rm eff}$ is the dimensionless nuclear matrix element (NME). The phase space factor is given by
\begin{equation}\label{Gij}
G_{ab}^{(0)} = \frac1{g_A^4 m_e^2}  \frac{2 \widetilde A^2}{3  \log2} \int_{m_e}^{E^{\rm max}_{e_1}}dE_{e_1}\int_{m_e}^{E^{\rm max}_{e_2}}dE_{e_2}
\int_{m_a}^{E_{a}^{\rm max}}dE_{a}  
 \Big(\langle K_{\mathcal N}\rangle^2 + \langle L_{\mathcal N}\rangle^2 + \langle K_{\mathcal N}\rangle\langle L_{\mathcal N}\rangle\Big)\,  f_{11}^{(0)}\,\omega_{ab} \,,
\end{equation}
where 
\begin{align}
\widetilde A &= \frac12 (Q_{\beta\beta}+2m_e) + \langle E_{\mathcal N}\rangle - E_I\,,\\
\langle K_{\mathcal N}\rangle &= \frac1{E_{e_1}+E_{a}+\langle E_{\mathcal N}\rangle - E_I} + \frac1{E_{e_2}+E_{b}+\langle E_{\mathcal N}\rangle - E_I} \,,\label{eqKN}\\ 
\langle L_{\mathcal N}\rangle &= \frac1{E_{e_1}+E_{b}+\langle E_{\mathcal N}\rangle - E_I} + \frac1{E_{e_2}+E_{a}+\langle E_{\mathcal N}\rangle - E_I}\,,\label{eqLN}
\end{align}
with $E_I$ the energy of the initial nucleus, $\langle E_{\mathcal N}\rangle$ a suitable excitation energy in the intermediate nucleus and
\begin{equation}\label{omega}
 \omega_{ab} = \frac{g_A^4 G_\beta^4}{64\pi^7}\, p_{a}E_{a} p_{b} E_{b} p_{e_1} E_{e_1} p_{e_2} E_{e_2}\,.
\end{equation}
Here, $E_X$ and $p_X=\sqrt{E_X^2-m_X^2}$ denote the energy and the modulus of the 3-momentum of the particle $X=e_1, e_2, a,b$ with mass $m_X$, and subject to the condition 
$E_{b}=Q_{\beta\beta}+2m_e-E_{e_1}-E_{e_2}-E_{a}$ from energy conservation. The integration limits read $E_{e_1}^{\rm max} = Q_{\beta\beta}+m_e-m_{a}-m_{b}$, $E_{e_2}^{\rm max} =Q_{\beta\beta}+2m_e-E_{e_1}-m_{a}-m_{b}$ and $E_{a}^{\rm max}=Q_{\beta\beta}+2m_e-E_{e_1}-E_{e_2}-m_{b}$, which for $m_a,m_b\neq 0$ shifts the end-point of the spectrum to lower values. Finally, the  factor $f_{11}^{(0)}$ originates from the Coulomb interaction of the electrons with the daughter nucleus, which we parameterize using the Fermi function~\cite{Doi:1985dx}
\begin{equation}\label{f110}
f_{11}^{(0)} \simeq F_0(Z+2,E_{e_1})\, F_0(Z+2,E_{e_2})\,.
\end{equation}
In the limit $a,b=\nu$ and $m_\nu=0$ we recover the results of~\cite{Kotila:2013gea}.

The complete analysis of the spectrum requires a numerical evaluation of Eq.~\eqref{dGammadT}. 
However, we also identified accurate analytical approximations for the dominant \nnbb\ and \nNbb\ channels (\NNbb\ is negligible for small $\sin\theta$), which assumes the Primakoff-Rosen approximation~\cite{Primakoff:1959chj} for the Fermi function and neglects the lepton energies in 
\begin{equation}
\langle K_{\mathcal N}\rangle \simeq\ \langle L_{\mathcal N}\rangle \simeq \frac4{2\widetilde A-Q_{\beta\beta}-2m_e} \approx\frac2{\widetilde A}\,.
\end{equation} 
Then, neglecting also the active neutrino mass but keeping the sterile neutrino one, we obtain:
\begin{equation}\label{dGanal}
\frac{dG_{ab}}{dT} \approx \frac{G_\beta^4 m_e^8}{7200\pi^7\log2}
\bigg[ \frac{2\pi\alpha(Z+2)}{1-\exp\big(-2\pi\alpha(Z+2)\big)}\bigg]^2\, F_{ab}(T)\,,
\end{equation}
where we have introduced the form factors
\begin{align}
F_{\nu\nu}(T)=&~T~(T^4+10T^3+40T^2+60T+30) ~ (T_0-T)^5\,,&\quad\label{eqFTnunu}\\
F_{\nu N}(T)=&\,\frac T2 (T^4+10T^3+40T^2+60T+30) \,
\bigg\{ \Big(2(T_0-T)^4-9x_N^2(T_0-T)^2-8x_N^4\Big)\sqrt{(T_0-T)^2-x_N^2} \nonumber\\
&+15 x_N^4(T_0-T)\Big[ \log\left(T_0-T+\sqrt{(T_0-T)^2-x_N^2}\right) - \log x_N\Big]\bigg\}\,.
\end{align}
\begin{figure*}[]
\begin{center}
\includegraphics[width=.49\textwidth]{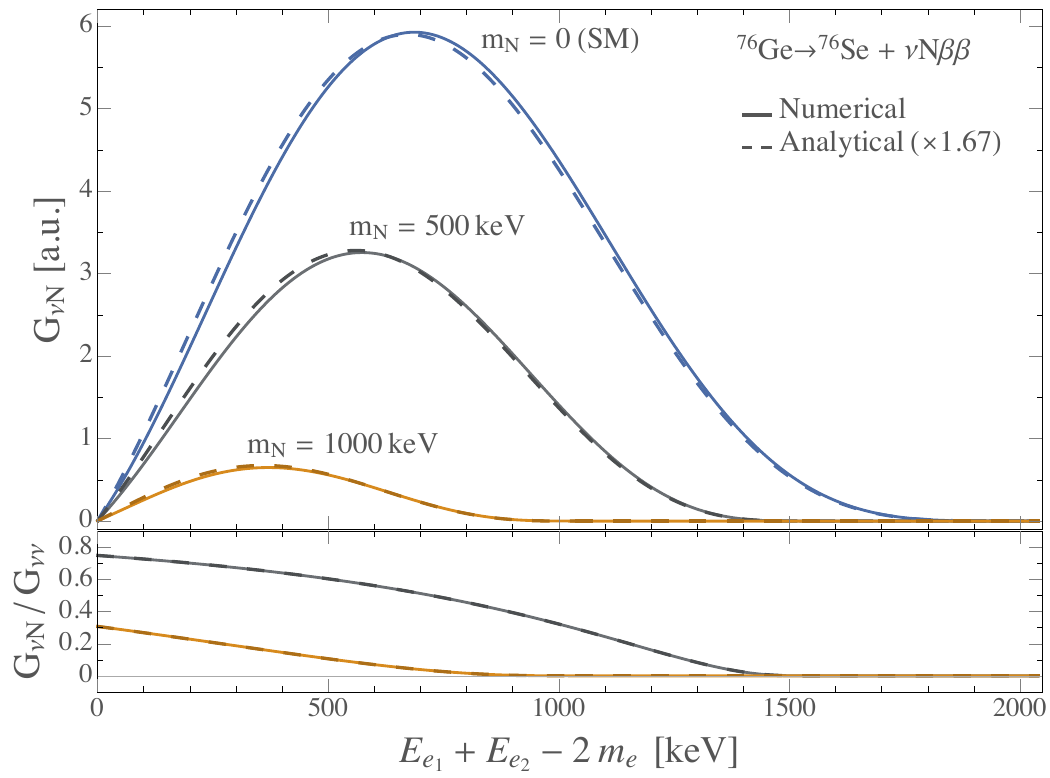}
\includegraphics[width=.49\textwidth]{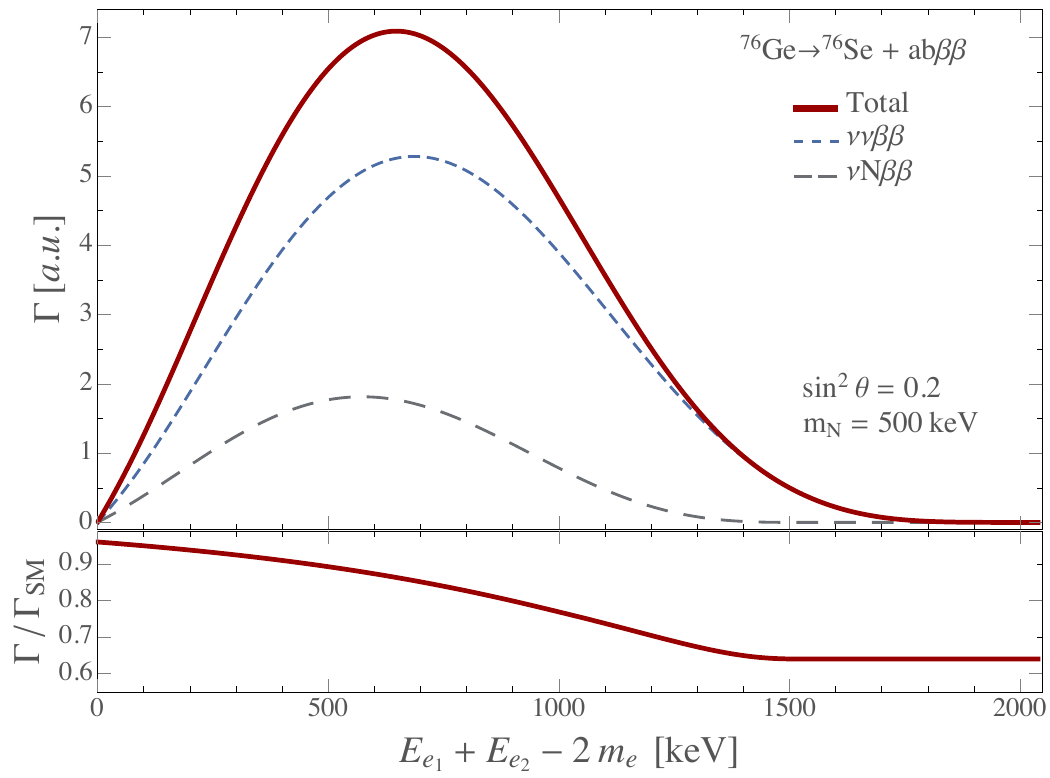}
\caption{Left: Phase space for different sterile neutrino mass hypotheses computed using a numerical evaluation of Eq.~\eqref{dGammadT} and the analytical approximation of Eq.~\eqref{dGanal}.
Right: \nnbb, \nNbb\ and summed energy spectra for $m_N=500$\,keV and an illustrative large mixing $\sin^2\theta=0.2$.}\label{spectrum1}
\end{center}
\end{figure*}

The left panel of Fig.~\ref{spectrum1} displays the phase space resulting from our numerical calculation applied to $^{76}$Ge ($Q_{\beta\beta}\simeq2039$\,keV), showing how the presence of a final massive invisible fermion modifies the energy spectrum, shifting the end-point as well as the peak to lower values as $m_N$ increases.
We can also see that the analytical expression reproduces the full result to a good approximation, up to a global factor of 1.67.
This factor is due to the approximation we considered, but it does not affect the analysis, which is sensitive to the $G_{\nu N}/G_{\nu\nu}$ ratio. 
We stress that, contrary to the single-$\beta$ decay spectrum, the end-point of the \nNbb\ spectrum is smooth and therefore the total spectrum does not manifest a kink at $Q_{\beta\beta}-m_N$. 
Nevertheless, the spectrum differs from the standard one, allowing \oobb\ experiments to probe this hypothesis by measuring the \nnbb\ spectrum very accurately.

\section{Double-$\pmb{\beta}$ decay into $\boldsymbol{Z_2}$-odd singlet fermions}\label{secZ2}

In general, the production of a pair of exotic particles is strongly suppressed compared to the production of a single one and can be neglected. However, there are scenarios in which the single production is forbidden and only the double production can take place. These scenarios cannot be tested by single-$\beta$ decay experiments, whereas \oobb\ decay experiments are still sensitive to the double production and have the unique opportunity to explore these channels.

In this section we consider a variant of the previous model, in which the symmetry group is extended by a discrete $Z_2$ symmetry, possibly related to the dark matter sector, which is exact or mildly broken in the electroweak vacuum.
We assume that all Standard Model particles are even under the $Z_2$ symmetry, while the neutral singlet fermion is odd. We will denote the $Z_2$-odd neutral singlet fermion as $\chi$ to differentiate it from the sterile neutrino, since the $Z_2$ symmetry forbids the mass term $m_D \overline{\nu} N$  characteristic of the latter. Correspondingly, the $Z_2$ symmetry forbids the single-$\beta$ decay $(A,Z)\rightarrow (A,Z+1)+e+\chi$,  as well as the double-$\beta$ decay $(A,Z)\rightarrow (A,Z+2)+2e+\nu+\chi$. However, the double-$\beta$ decay $(A,Z)\rightarrow(A,Z+2)+2e+2\chi$ is possible (\XXbb\ decay). The search for spectral distortions in the double-$\beta$ decay spectrum can therefore probe scenarios with light $Z_2$-odd singlet fermions.

The total differential decay rate receives in this scenario two contributions:
\begin{align}
\frac{d\Gamma}{dT} =&~  \frac{d\Gamma_{\nu\nu}}{dT}\,  \theta(T_0-T)  +\frac{d\Gamma_{\chi\chi}}{dT} \, \theta(T_0-T-2x_\chi)\,, 
\end{align}
where  $d\Gamma_{\nu\nu}/dT$ is the Standard Model contribution, defined in Eq.~\eqref{dGammadT}, and 
$d\Gamma_{\chi\chi}/dT$ is the exotic contribution. Here, $T$ and $T_0$ are defined as in  Eq.~(\ref{eq:dGammadT}), with $x_\chi=m_\chi/m_e$. 
The exotic contribution is very model dependent. For definiteness, we will consider the following effective interaction between the active neutrinos and $\chi$:
\begin{equation}\label{Lagchi}
\mathcal L = g_\chi \nu \nu\chi \chi\,, 
\end{equation}
with a constant coupling $g_\chi$. We use the results of~\cite{Deppisch:2020sqh}, that considered a similar four-fermion scalar interaction as in Eq.~\eqref{Lagchi} but for neutrino self-interactions, to relate the decay rate for \XXbb\ to that of \NNbb.
We obtain
\begin{equation}\label{Ratechi}
\frac{d\Gamma_{\chi\chi}}{dT} = \frac{g_\chi^2 m_e^2}{8\pi^2 R^2} |\mathcal M_{0\nu}|^2  \frac{dG_{NN}^{(0)}}{dT}\,,
\end{equation}
where $\mathcal M_{0\nu}$ is the NME of the \oobb\ process, $R$ is the nuclear radius and the phase factor $G_{NN}^{(0)}$ is given in Eq.~\eqref{Gij} (replacing $m_N$ by $m_\chi$).
The translation of experimental measurements into constraints on $g_\chi$ will hence need to rely on NME calculations.

\section{Discovery potential of future $\boldsymbol{0\nu}\pmb{\beta\beta}$ decay experiments}\label{Projections}
A large experimental program has been mounted to search for the \oobb\ decay of different isotopes and using different detection techniques~\cite{Dolinski:2019nrj}. The current-generation experiments set the ground for the identification of the best techniques for further investment. The next-generation experiments are currently being proposed, designed, or in construction. Among the experiments proposed for the next decade, we will focus on LEGEND~\cite{Abgrall:2017syy}, nEXO~\cite{Kharusi:2018eqi} and CUPID~\cite{CUPIDInterestGroup:2019inu}. We selected these experiments because their efficiencies and uncertainties in the search for massive fermions can be inferred from the \nnbb\ analysis published by their predecessors, {\it i.e.}, GERDA~\cite{Agostini:2015nwa}, EXO-200~\cite{Albert:2013gpz} and CUPID-Mo~\cite{Armengaud:2019rll}. The parameters assumed for each experiment are listed in Tab.~\ref{tab:exp}.
\begin{table}[]
    \centering
    \setlength{\tabcolsep}{6pt}
    \renewcommand{\arraystretch}{1.3}

    \begin{tabular}{ccccc}
        \hline
            Isotope & Experiment & $T_{1/2}^{\nnbb}$ [yr] & Efficiency & Exposure [mol yr] \\
        \hline 
             $^{76}$Ge 
             & GERDA\,/\,LEGEND-200\,/\,LEGEND-1000
             & $2.0\cdot10^{21}$~\cite{Abramov:2019hhh} 
             & 75\%~\cite{Agostini:2015nwa}
             & $1.4\cdot{10^3}$~\cite{Agostini:2020xta}\,/\,$1.4\cdot{10^4}$~\cite{Abgrall:2017syy}\,/\,$1.4\cdot{10^5}$~\cite{Abgrall:2017syy}
             \\
             $^{136}$Xe & EXO-200\,/\,nEXO 
             & $2.2\cdot10^{21}$~\cite{Albert:2013gpz}
             & 85\%~\cite{Albert:2013gpz}
             & $1.7\cdot{10^3}$~\cite{Anton:2019wmi}\,/\,$3.7\cdot{10^5}$~\cite{Kharusi:2018eqi} 
             \\
             $^{100}$Mo & CUPID-Mo\,/\,CUPID
             & $7.1\cdot10^{18}$~\cite{Armengaud:2019rll}
             & 91\%~\cite{Armengaud:2019rll}
             & $6.5\cdot{10^{-1}}$~\cite{Armengaud:2019rll}\,/\,$2.7\cdot{10^4}$~\cite{CUPIDInterestGroup:2019inu} 
             \\

        \hline
    \end{tabular}
    \caption{Parameters used for the sensitivity projections of each experiment. $T_{1/2}^{2\nu\beta\beta}$ is the double-$\beta$ decay half-life, {\it i.e.} the inverse of the decay rate $R_{2\nu\beta\beta}$. The detection efficiency refers to fraction of \nnbb\ decay events that populate the energy window of interest after all analysis cuts. The exposure is given in number of moles of detector material per year of live time.}
    \label{tab:exp}
\end{table}
 
While the target isotope and the backgrounds vary across these experiments, the analysis to search for light exotic fermions is always conceptually the same. The energy window of interest goes from the detector threshold to the Q-value of the decay. In this window, the majority of the events is due to \nnbb\ decays ($>95$\% in the current-generation experiments). The other events can be due to a multitude of processes, for instance, natural radioactivity and cosmic rays. The most important parameters affecting the sensitivity of an experiment to light exotic fermions are the exposure, the background rate and the systematic uncertainties due to the energy reconstruction. The exposure is given by the product of the number of observed nuclei and the observation time. The background rate is primarily given by the \nnbb\ decay rate with a subdominant contribution due to the other sources. The systematic uncertainties related to the event energy reconstruction can largely differ between experiments, but in general, their impact can be parameterized through an energy-dependent shape factor.

To accurately quantify the sensitivity of the experiments, we have implemented a comprehensive frequentist analysis framework. Distortions of the double-$\beta$ decay energy distribution due to light exotic fermions are searched through a binned maximum-likelihood fit based on a profile-likelihood test statistic~\cite{Cowan:2010js}. Each process possibly contributing to the count rate in the energy window of interest should be added to the fit through a probability distribution function (PDF) that describes its expected event energy distribution. For this work, we use a PDF for the sought-after signal and one for the dominant \nnbb\ decay. Both these PDFs are based on the calculations described in the previous sections. We use a third uniform probability distribution to account for other generic sub-dominant background sources. The actual shape of this third PDF affects only marginally our results. The parameters of the fit are the scaling factors of each PDF, {\it i.e.} the number of events attributed to each process.

The probability distributions of the test statistic are computed from large ensembles of pseudo-data generated under different hypotheses on the signal rate. This approach provides the right coverage by construction, including when the signal rate is close to the physical border.

Systematic uncertainties can bias the event energy reconstruction. Many detector-specific sources of bias should be considered. However, their overall impact can be parameterized through a shape factor with the form $f(E)=1+a\cdot E+b\cdot E^2+c/E$ where $a$, $b$ and $c$ are parameters that are considered to be known with limited accuracy ({\it i.e.} $\sigma_a$, $\sigma_b$ and $\sigma_c$). Current-generation experiments are typically able to control the energy reconstruction bias at the per-cent level. To incorporate this systematic uncertainty in our analysis, we randomize $a$, $b$ and $c$ during the generation of the pseudo-data and run the fit using not-deformed PDFs~\cite{Cousins:1991qz}. The parameters are sampled from normal distributions centred at 0 and with a sigma of $10^{-3}\,\text{keV}^{-1}$, $10^{-6}\,\text{keV}^{-2}$ and $10^{-3}\,\text{keV}$, respectively. We tested that this parameterization covers the maximal distortions estimated by the experiments. The result of this procedure is a broadening of the test statistic distribution and a reduction of the power of the statistical analysis. 

\begin{figure}[]
    \begin{center}
        \includegraphics[width=.49\textwidth]{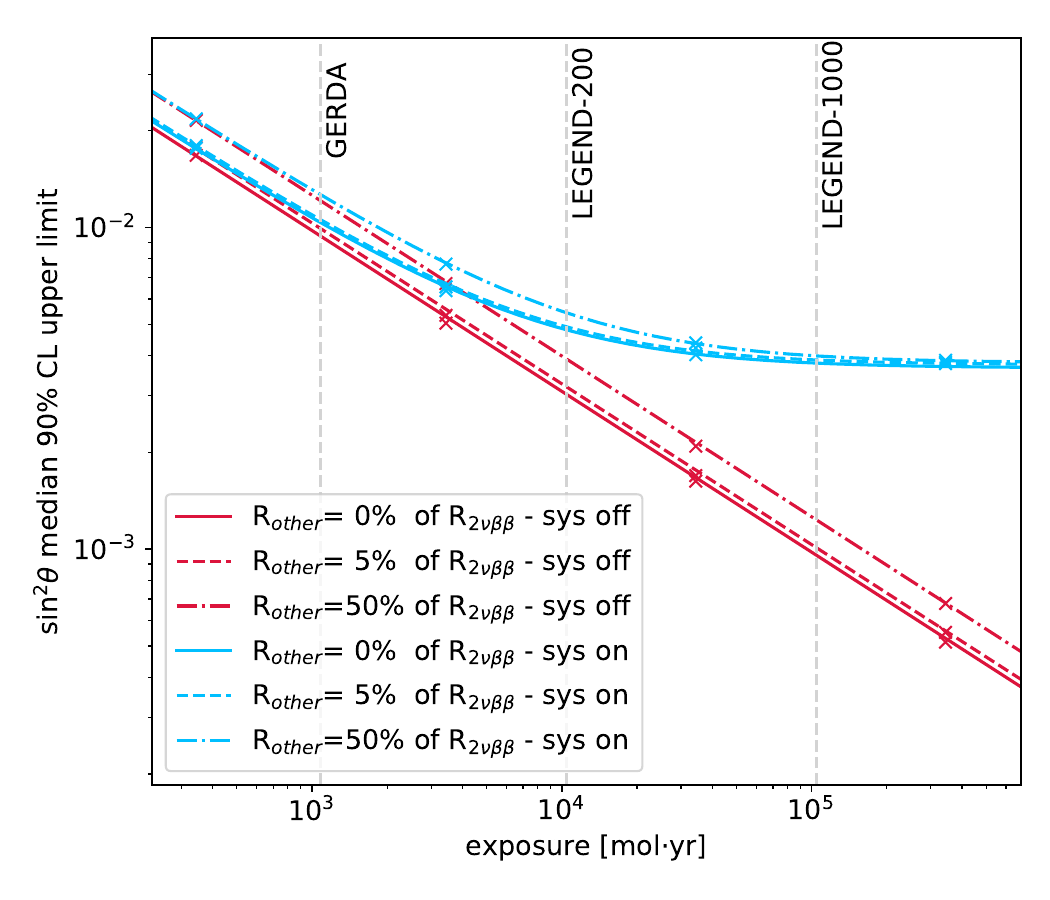}
        \caption{Sensitivity of a $^{76}$Ge double-$\beta$ decay experiment searching for massive sterile neutrinos with a mass of 500\,keV, given as a function of exposure and background level. The sensitivity is expressed in terms of the median 90\% C.L. upper limit on the squared of the mixing angle ($\sin^2\theta$), computed assuming no signal. The markers indicate the value computed using a full frequentist analysis, while the lines show their best fit with a function $f(\mathcal{E})=\sqrt{\sigma_{\text{stat}}^2(\mathcal{E}) + \sigma_{\text{sys}}^2} = \sqrt{\alpha/\mathcal{E}+\beta}$ where $\alpha$ and $\beta$ are free parameters and $\mathcal{E}$ is the exposure. The plot shows the impact of additional backgrounds with a rate $R_{\rm other}$ which is expressed w.r.t. the irreducible \nnbb\ decay rate $R_{\nnbb}$.}
        \label{sensitivity_exposure}
    \end{center}
\end{figure}

Figure~\ref{sensitivity_exposure} shows our projected sensitivity for a generic $^{76}$Ge experiment and a sterile neutrino with a mass of 500\,keV. The sensitivity for different background rates and exposure values is given in terms of the median 90\% confidence level (C.L.) upper limit on $\sin^2\theta$~\cite{Cowan:2010js}.
The upper limit scales approximately as the sum in quadrature of the statistical and systematic uncertainties. 
As long as the statistical uncertainty is dominant, the upper limit improves by increasing the exposure. The sensitive saturates when the statistical uncertainty becomes comparable with the systematic one.
The vertical lines in the figure mark the exposure that has been collected by GERDA, as well as the exposure planned for the two phases of LEGEND. LEGEND has the potential to improve the sensitivity by an order of magnitude compared to GERDA, assuming that the systematic uncertainties can be kept below the statistical uncertainty. A significant improvement is expected even in the conservative scenario in which the systematic uncertainties are not reduced with respect to the current level. 

The parameter of interest for sterile neutrino searches is the mixing angle, i.e., the ratio between the number of reconstructed \nNbb\ and \nnbb\ decay events. If we express the statistical uncertainty on the number of decays in terms of the \nnbb\ decay rate ($R_{\nnbb}$) and exposure ($\mathcal{E}$) and propagate these uncertainties to the mixing angle, we find that the sensitivity scales as $\sin^2\theta\propto\sqrt{1/(\mathcal{E}\cdot R_{\nnbb})}$. The parameter of interest for the search of  $Z_2$-odd fermions is instead the number of \XXbb\ events and its conversion into a a coupling constant requires also the NME nuclear matrix element $\mathcal{M}$ and phase space factor $G$.  By propagating the uncertainties, we obtain that the sensitivity scales as $g_{\chi}^2\propto G^{-1} \mathcal{M}^{-2}\sqrt{R_{\nnbb}/\mathcal{E}}$. In both cases the sensitivity is proportional to $1/\sqrt{\mathcal{E}}$, but the functional dependence from $R_{\nnbb}$ is inverted. In sterile neutrino searches, the larger \nnbb\ decay rate, the larger the number of \nNbb\ decays. An increase of $R_{\nnbb}$ will hence lead to a reduction of the statistical uncertainty. In the search for \XXbb\ decays, the \nnbb\ decays constitute a background. The higher the background, the lower the sensitivity is. Because of this difference, isotopes which are favorable for one search might not be optimal for the other.

\begin{figure}[]
\begin{center}
\includegraphics[width=0.49\textwidth]{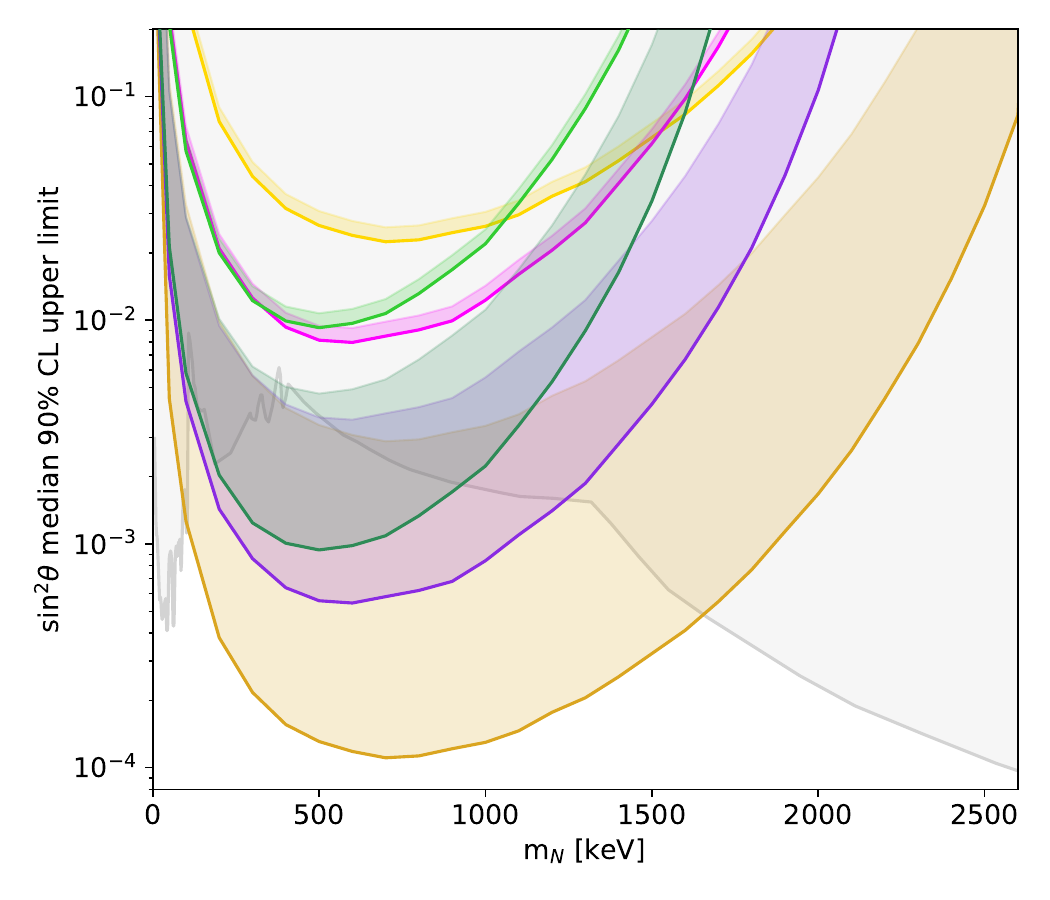}
\includegraphics[width=0.49\textwidth]{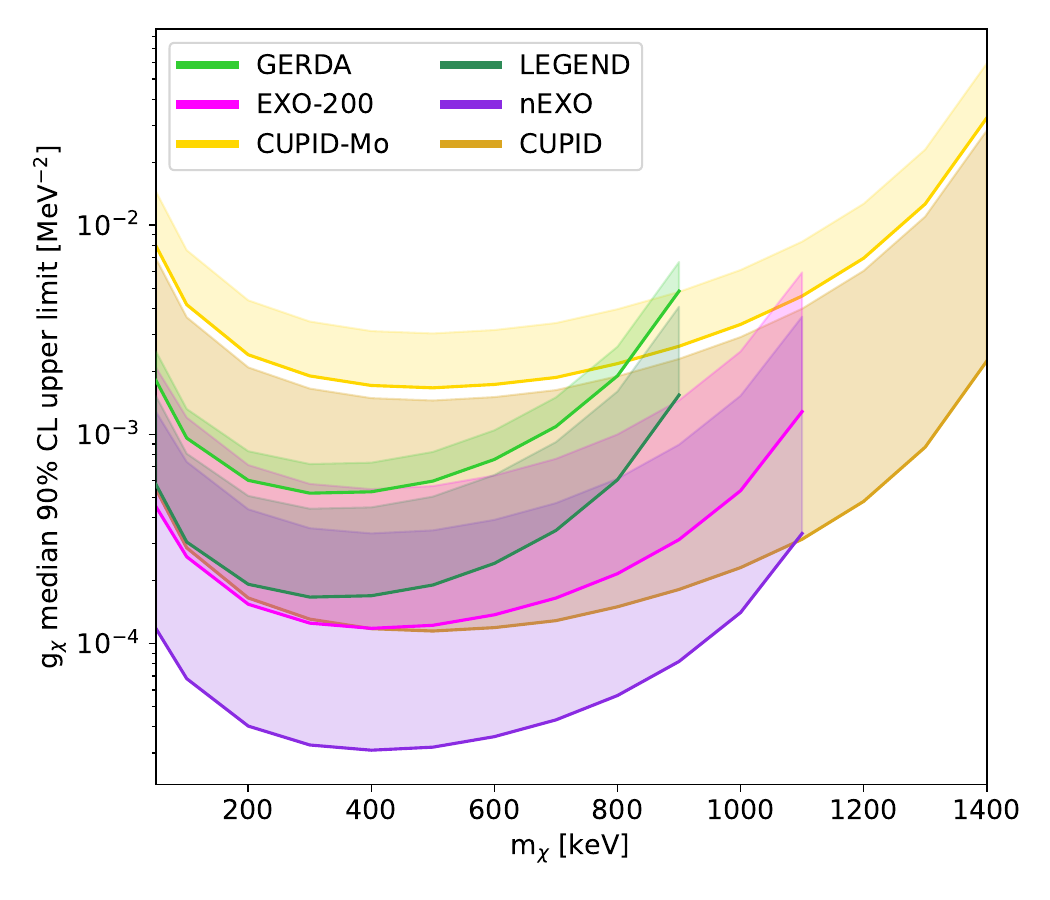}
\caption{Sensitivity to light exotic fermions for a selection of double-$\beta$ decay experiments. The left panel shows the median 90\% C.L. upper limit on the squared of the sterile neutrino mixing angle as a function of its mass.  The experimental constraints are displayed through a band covering a range of sensitivities that goes from the most optimistic scenario (systematic uncertainty smaller than statistical uncertainty), till a conservative scenario (systematic uncertainty at the level of past analyses). Existing sterile neutrino constraints from single-$\beta$ decay experiments~\cite{Holzschuh:2000nj,Deutsch:1990ut,Schreckenbach:1983cg,Derbin:2018dbu} and solar neutrinos~\cite{Bellini:2013uui} are shown in the background.
The right panel shows the median upper limit on the coupling constant between $Z_2$-odd fermion and the neutrinos assuming the effective interaction in Eq.~\eqref{Lagchi}. The spread of the bands account for the systematic uncertainties as in the case of the sterile neutrinos, but additionally it covers also the full range of possible NME values found in literature~\cite{Barea:2015kwa,Menendez:2017fdf,Horoi:2015tkc,Vaquero:2014dna,Rodriguez:2010mn,Song:2017ktj,Hyvarinen:2015bda,Mustonen:2013zu,Fang:2018tui,Simkovic:2018hiq,Coraggio:2020hwx}.}\label{sensitivity_mass}
\end{center}
\end{figure}

Our sensitivity projections for the search of sterile neutrinos and of the $Z_2$-odd fermions are shown in Fig.~\ref{sensitivity_mass}. The sensitivities are shown using bands in which the upper edge corresponds to the conservative scenario in which the systematic uncertainties will not be improved compared to the current-generation experiments, while the lower edge corresponds to the optimistic scenario in which they will be sub-dominant. These two scenarios define the ballpark for the sensitivity of future experiment and the results for intermediate scenarios can be interpolated from these two cases. The sensitivity evolution as a function of the fermion mass has a parabolic shape. Its minimum depends on the Q-value of the decay and corresponds to the fermion mass for which the experiment is most sensitive. The experiments quickly loose sensitivity towards vanishing masses. This is because the smaller is the fermion mass, the smaller is the spectral distortion. A similar loss of sensitivity occurs at larger masses, where the fermion mass approaches the maximum energy available in the decay and the phase space shrinks\footnote{Fermions with masses up to the Q-value of the decay could be produced, although with very suppressed phase space. When the simultaneous production of two fermions is considered, the maximum mass is equal to a half of the Q-value.}. The exposure, isotope properties, and experimental parameters define the offset of the curves. These projections are done assuming that each experiment can extend its analysis window down to arbitrarily small energies. If this is not the case, the upper edge of the curve will be lowered by the value of the energy threshold and the offset will slightly increase because of the reduction in detection efficiency.
The sensitivity on the coupling constant $g_\chi$ is computed from the decay rate of \XXbb\, using Eq.~\eqref{Ratechi}. In addition to the systematic uncertainty considered for the sterile neutrino search, the width of the bands accounts also for the uncertainties due to the  NME calculations.

\section{Results and Outlook}\label{Conclusions} 

The main results of our analysis are displayed in Fig.~\ref{sensitivity_mass}.
The left panel shows the projected upper limits on the active-sterile mixing, parametrized as $\sin^2\theta$, as a function of the sterile neutrino mass. We also show in the plot the current limits from single-$\beta$ decay experiments~\cite{Holzschuh:2000nj,Deutsch:1990ut,Schreckenbach:1983cg,Derbin:2018dbu} and from solar neutrinos~\cite{Bellini:2013uui}. As can be seen from the plot, the sensitivity of current double-$\beta$ decay experiments is weaker than the existing limits, but only by a factor of a few. The larger exposure of future double-$\beta$ decay experiments encourages a dedicated search for these exotic decay channels. 

Indeed, our sensitivity study demonstrates the potential of future double-$\beta$ decay experiments to improve the current limits on the active-sterile mixing angle in this mass range. Fig.~\ref{sensitivity_mass} shows our projected sensitivities for LEGEND-1000, nEXO and CUPID, assuming no improvement in the systematic uncertainties with respect to the current-generation experiments (upper side of the bands), or assuming that the systematic uncertainties will be reduced below the statistical ones (lower side of the bands). In the most conservative scenario, the sensitivity of future searches will be comparable to the current limits. However, even with a modest reduction of the systematic uncertainties, next-generation experiments will explore uncharted regions of the sterile neutrino parameter space, even reaching  $\sin^2\theta\sim10^{-3}-10^{-4}$ for $m_N\sim(100-1600)$\,keV.

Double-$\beta$ decay experiments have also the capability of probing models in which only the double production of light exotic fermions is allowed. This previously overlooked opportunity can lead to the first constraints on this kind of models.
This is shown in the right panel of Fig.~\ref{sensitivity_mass} for the $Z_2$-odd singlet $\chi$ introduced in Sec.~\ref{secZ2}.
The sensitivities of current experiments for the effective coupling $g_\chi$ lie between $10^{-2}$ and $10^{-3}~{\rm MeV}^{-2}$, which could be improved up to $10^{-4}~{\rm MeV}^{-2}$ with a favorable NME value and negligible systematic uncertainties. 

Assuming that this effective interaction is originated at some scale $\Lambda$ with $\mathcal O(1)$ dimensionless Wilson coefficient, this would imply that $\Lambda\gtrsim 100$\,MeV. This is comparable to the typical momentum transfer in double-$\beta$ decay experiments, and therefore the effective field theory approach employed in this work to recast the limits on the decay rate into limits on the coupling strength should be taken with a grain of salt. Yet, the search for distortions in the double-$\beta$ decay spectrum due to the emission of two light exotic particles (fermions or scalars) is well motivated theoretically and deserves further investigation. 
\\

Cosmological observations offer complementary constraints on the existence of long-lived light exotic fermions. For the values of the mixing angles that can be probed by future and planned double-$\beta$ decay experiments, the light exotic fermion would thermalize with the primeval plasma, and thereby alter the successful predictions of the standard Big Bang nucleosynthesis scenario~\cite{Vincent:2014rja,Sabti:2019mhn}. One should note that the strong constraints from cosmology can be circumvented if the light exotic fermion decays before the onset of the nucleosynthesis reactions. This is the case, for instance, if the $Z_2$ symmetry is only approximate, so that $\chi$ is still produced in pairs and stable within the detectors, but is not cosmologically long-lived, or if $N$ decays promptly into invisible particles. It is therefore of utmost importance to perform  searches for exotic light fermions in laboratory experiments, as their implications for cosmology depend on additional parameters (most notably their lifetime).

 To conclude, in this letter we have proposed and explored the possibility of searching for the production of light exotic fermions using double-$\beta$ decay experiments. We have derived the double-$\beta$ decay spectrum when one or two light exotic fermions are emitted in the final state. We have also studied how these new channels can be searched for through a fit, and we have estimated the projected sensitivity  for various future experiments considering the impact of both statistical and systematic uncertainties. We have found that double-$\beta$ decay experiments constitute an promising avenue (and for some scenarios unique) to search for new physics. We therefore encourage the experimental collaborations to perform dedicated searches for this class of scenarios, properly including all the systematic and statistical uncertainties.

~\\
{\it Note added.} During the final stages of this project, a preprint by Bolton {\it et al.}~\cite{Bolton:2020ncv} appeared, also exploring the possibility of searching for sterile neutrinos with double-$\beta$ decay experiments. While they focus exclusively on sterile neutrinos (left and right-handed models) we considered generic single and double production of light exotic fermions. Their procedure to derive the sensitivity projections and their treatment of the systematic uncertainties differ from ours. Nevertheless, in the aspects where our analysis overlap, the results are qualitatively similar.

\begin{acknowledgements}
X.M. would like to thank Claudia Garc\'ia-Garc\'ia for useful discussions.
This work has been supported by the Collaborative Research Center SFB1258, by the Deutsche Forschungsgemeinschaft (DFG, German Research Foundation) under Germany's Excellence Strategy - EXC-2094 - 390783311, and by the Science and Technology Facilities Council [grant number ST/T004169/1].
X.M. is supported by the Alexander von Humboldt Foundation. 
\end{acknowledgements}

\bibliographystyle{apsrev4-1}
\bibliography{biblio}

\begin{thebibliography}{52}%
\makeatletter
\providecommand \@ifxundefined [1]{%
 \@ifx{#1\undefined}
}%
\providecommand \@ifnum [1]{%
 \ifnum #1\expandafter \@firstoftwo
 \else \expandafter \@secondoftwo
 \fi
}%
\providecommand \@ifx [1]{%
 \ifx #1\expandafter \@firstoftwo
 \else \expandafter \@secondoftwo
 \fi
}%
\providecommand \natexlab [1]{#1}%
\providecommand \enquote  [1]{``#1''}%
\providecommand \bibnamefont  [1]{#1}%
\providecommand \bibfnamefont [1]{#1}%
\providecommand \citenamefont [1]{#1}%
\providecommand \href@noop [0]{\@secondoftwo}%
\providecommand \href [0]{\begingroup \@sanitize@url \@href}%
\providecommand \@href[1]{\@@startlink{#1}\@@href}%
\providecommand \@@href[1]{\endgroup#1\@@endlink}%
\providecommand \@sanitize@url [0]{\catcode `\\12\catcode `\$12\catcode
  `\&12\catcode `\#12\catcode `\^12\catcode `\_12\catcode `\%12\relax}%
\providecommand \@@startlink[1]{}%
\providecommand \@@endlink[0]{}%
\providecommand \url  [0]{\begingroup\@sanitize@url \@url }%
\providecommand \@url [1]{\endgroup\@href {#1}{\urlprefix }}%
\providecommand \urlprefix  [0]{URL }%
\providecommand \Eprint [0]{\href }%
\providecommand \doibase [0]{http://dx.doi.org/}%
\providecommand \selectlanguage [0]{\@gobble}%
\providecommand \bibinfo  [0]{\@secondoftwo}%
\providecommand \bibfield  [0]{\@secondoftwo}%
\providecommand \translation [1]{[#1]}%
\providecommand \BibitemOpen [0]{}%
\providecommand \bibitemStop [0]{}%
\providecommand \bibitemNoStop [0]{.\EOS\space}%
\providecommand \EOS [0]{\spacefactor3000\relax}%
\providecommand \BibitemShut  [1]{\csname bibitem#1\endcsname}%
\let\auto@bib@innerbib\@empty
\bibitem [{\citenamefont {Bolton}\ \emph
  {et~al.}(2020{\natexlab{a}})\citenamefont {Bolton}, \citenamefont
  {Deppisch},\ and\ \citenamefont {Bhupal~Dev}}]{Bolton:2019pcu}%
  \BibitemOpen
  \bibfield  {author} {\bibinfo {author} {\bibfnamefont {P.~D.}\ \bibnamefont
  {Bolton}}, \bibinfo {author} {\bibfnamefont {F.~F.}\ \bibnamefont
  {Deppisch}}, \ and\ \bibinfo {author} {\bibfnamefont {P.}~\bibnamefont
  {Bhupal~Dev}},\ }\href {\doibase 10.1007/JHEP03(2020)170} {\bibfield
  {journal} {\bibinfo  {journal} {JHEP}\ }\textbf {\bibinfo {volume} {03}},\
  \bibinfo {pages} {170} (\bibinfo {year} {2020}{\natexlab{a}})},\ \Eprint
  {http://arxiv.org/abs/1912.03058} {arXiv:1912.03058 [hep-ph]} \BibitemShut
  {NoStop}%
\bibitem [{\citenamefont {Deutsch}\ \emph {et~al.}(1990)\citenamefont
  {Deutsch}, \citenamefont {Lebrun},\ and\ \citenamefont
  {Prieels}}]{Deutsch:1990ut}%
  \BibitemOpen
  \bibfield  {author} {\bibinfo {author} {\bibfnamefont {J.}~\bibnamefont
  {Deutsch}}, \bibinfo {author} {\bibfnamefont {M.}~\bibnamefont {Lebrun}}, \
  and\ \bibinfo {author} {\bibfnamefont {R.}~\bibnamefont {Prieels}},\ }\href
  {\doibase 10.1016/0375-9474(90)90541-S} {\bibfield  {journal} {\bibinfo
  {journal} {Nucl. Phys. A}\ }\textbf {\bibinfo {volume} {518}},\ \bibinfo
  {pages} {149} (\bibinfo {year} {1990})}\BibitemShut {NoStop}%
\bibitem [{\citenamefont {Schreckenbach}\ \emph {et~al.}(1983)\citenamefont
  {Schreckenbach}, \citenamefont {Colvin},\ and\ \citenamefont
  {Von~Feilitzsch}}]{Schreckenbach:1983cg}%
  \BibitemOpen
  \bibfield  {author} {\bibinfo {author} {\bibfnamefont {K.}~\bibnamefont
  {Schreckenbach}}, \bibinfo {author} {\bibfnamefont {G.}~\bibnamefont
  {Colvin}}, \ and\ \bibinfo {author} {\bibfnamefont {F.}~\bibnamefont
  {Von~Feilitzsch}},\ }\href {\doibase 10.1016/0370-2693(83)90858-4} {\bibfield
   {journal} {\bibinfo  {journal} {Phys. Lett. B}\ }\textbf {\bibinfo {volume}
  {129}},\ \bibinfo {pages} {265} (\bibinfo {year} {1983})}\BibitemShut
  {NoStop}%
\bibitem [{\citenamefont {Dolinski}\ \emph {et~al.}(2019)\citenamefont
  {Dolinski}, \citenamefont {Poon},\ and\ \citenamefont
  {Rodejohann}}]{Dolinski:2019nrj}%
  \BibitemOpen
  \bibfield  {author} {\bibinfo {author} {\bibfnamefont {M.~J.}\ \bibnamefont
  {Dolinski}}, \bibinfo {author} {\bibfnamefont {A.~W.}\ \bibnamefont {Poon}},
  \ and\ \bibinfo {author} {\bibfnamefont {W.}~\bibnamefont {Rodejohann}},\
  }\href {\doibase 10.1146/annurev-nucl-101918-023407} {\bibfield  {journal}
  {\bibinfo  {journal} {Ann. Rev. Nucl. Part. Sci.}\ }\textbf {\bibinfo
  {volume} {69}},\ \bibinfo {pages} {219} (\bibinfo {year} {2019})},\ \Eprint
  {http://arxiv.org/abs/1902.04097} {arXiv:1902.04097 [nucl-ex]} \BibitemShut
  {NoStop}%
\bibitem [{\citenamefont {Furry}(1939)}]{Furry:1939qr}%
  \BibitemOpen
  \bibfield  {author} {\bibinfo {author} {\bibfnamefont {W.}~\bibnamefont
  {Furry}},\ }\href {\doibase 10.1103/PhysRev.56.1184} {\bibfield  {journal}
  {\bibinfo  {journal} {Phys. Rev.}\ }\textbf {\bibinfo {volume} {56}},\
  \bibinfo {pages} {1184} (\bibinfo {year} {1939})}\BibitemShut {NoStop}%
\bibitem [{\citenamefont {Schechter}\ and\ \citenamefont
  {Valle}(1982)}]{Schechter:1981bd}%
  \BibitemOpen
  \bibfield  {author} {\bibinfo {author} {\bibfnamefont {J.}~\bibnamefont
  {Schechter}}\ and\ \bibinfo {author} {\bibfnamefont {J.}~\bibnamefont
  {Valle}},\ }\href {\doibase 10.1103/PhysRevD.25.2951} {\bibfield  {journal}
  {\bibinfo  {journal} {Phys. Rev. D}\ }\textbf {\bibinfo {volume} {25}},\
  \bibinfo {pages} {2951} (\bibinfo {year} {1982})}\BibitemShut {NoStop}%
\bibitem [{\citenamefont {Hirsch}\ \emph {et~al.}(1996)\citenamefont {Hirsch},
  \citenamefont {Klapdor-Kleingrothaus}, \citenamefont {Kovalenko},\ and\
  \citenamefont {Pas}}]{Hirsch:1995in}%
  \BibitemOpen
  \bibfield  {author} {\bibinfo {author} {\bibfnamefont {M.}~\bibnamefont
  {Hirsch}}, \bibinfo {author} {\bibfnamefont {H.}~\bibnamefont
  {Klapdor-Kleingrothaus}}, \bibinfo {author} {\bibfnamefont {S.}~\bibnamefont
  {Kovalenko}}, \ and\ \bibinfo {author} {\bibfnamefont {H.}~\bibnamefont
  {Pas}},\ }\href {\doibase 10.1016/0370-2693(96)00038-X} {\bibfield  {journal}
  {\bibinfo  {journal} {Phys. Lett. B}\ }\textbf {\bibinfo {volume} {372}},\
  \bibinfo {pages} {8} (\bibinfo {year} {1996})},\ \Eprint
  {http://arxiv.org/abs/hep-ph/9511227} {arXiv:hep-ph/9511227} \BibitemShut
  {NoStop}%
\bibitem [{\citenamefont {Díaz}(2014)}]{D_az_2014}%
  \BibitemOpen
  \bibfield  {author} {\bibinfo {author} {\bibfnamefont {J.~S.}\ \bibnamefont
  {Díaz}},\ }\href {\doibase 10.1103/physrevd.89.036002} {\bibfield  {journal}
  {\bibinfo  {journal} {Physical Review D}\ }\textbf {\bibinfo {volume} {89}}
  (\bibinfo {year} {2014}),\ 10.1103/physrevd.89.036002}\BibitemShut {NoStop}%
\bibitem [{\citenamefont {Deppisch}\ \emph
  {et~al.}(2020{\natexlab{a}})\citenamefont {Deppisch}, \citenamefont {Graf},\
  and\ \citenamefont {\v{S}imkovic}}]{Deppisch:2020mxv}%
  \BibitemOpen
  \bibfield  {author} {\bibinfo {author} {\bibfnamefont {F.~F.}\ \bibnamefont
  {Deppisch}}, \bibinfo {author} {\bibfnamefont {L.}~\bibnamefont {Graf}}, \
  and\ \bibinfo {author} {\bibfnamefont {F.}~\bibnamefont {\v{S}imkovic}},\
  }\href {\doibase 10.1103/PhysRevLett.125.171801} {\bibfield  {journal}
  {\bibinfo  {journal} {Phys. Rev. Lett.}\ }\textbf {\bibinfo {volume} {125}},\
  \bibinfo {pages} {171801} (\bibinfo {year} {2020}{\natexlab{a}})},\ \Eprint
  {http://arxiv.org/abs/2003.11836} {arXiv:2003.11836 [hep-ph]} \BibitemShut
  {NoStop}%
\bibitem [{\citenamefont {Deppisch}\ \emph
  {et~al.}(2020{\natexlab{b}})\citenamefont {Deppisch}, \citenamefont {Graf},
  \citenamefont {Rodejohann},\ and\ \citenamefont {Xu}}]{Deppisch:2020sqh}%
  \BibitemOpen
  \bibfield  {author} {\bibinfo {author} {\bibfnamefont {F.~F.}\ \bibnamefont
  {Deppisch}}, \bibinfo {author} {\bibfnamefont {L.}~\bibnamefont {Graf}},
  \bibinfo {author} {\bibfnamefont {W.}~\bibnamefont {Rodejohann}}, \ and\
  \bibinfo {author} {\bibfnamefont {X.-J.}\ \bibnamefont {Xu}},\ }\href
  {\doibase 10.1103/PhysRevD.102.051701} {\bibfield  {journal} {\bibinfo
  {journal} {Phys. Rev. D}\ }\textbf {\bibinfo {volume} {102}},\ \bibinfo
  {pages} {051701} (\bibinfo {year} {2020}{\natexlab{b}})},\ \Eprint
  {http://arxiv.org/abs/2004.11919} {arXiv:2004.11919 [hep-ph]} \BibitemShut
  {NoStop}%
\bibitem [{\citenamefont {Shrock}(1980)}]{Shrock:1980vy}%
  \BibitemOpen
  \bibfield  {author} {\bibinfo {author} {\bibfnamefont {R.}~\bibnamefont
  {Shrock}},\ }\href {\doibase 10.1016/0370-2693(80)90235-X} {\bibfield
  {journal} {\bibinfo  {journal} {Phys. Lett. B}\ }\textbf {\bibinfo {volume}
  {96}},\ \bibinfo {pages} {159} (\bibinfo {year} {1980})}\BibitemShut
  {NoStop}%
\bibitem [{\citenamefont {Riis}\ and\ \citenamefont
  {Hannestad}(2011)}]{Riis:2010zm}%
  \BibitemOpen
  \bibfield  {author} {\bibinfo {author} {\bibfnamefont {A.~S.}\ \bibnamefont
  {Riis}}\ and\ \bibinfo {author} {\bibfnamefont {S.}~\bibnamefont
  {Hannestad}},\ }\href {\doibase 10.1088/1475-7516/2011/02/011} {\bibfield
  {journal} {\bibinfo  {journal} {JCAP}\ }\textbf {\bibinfo {volume} {02}},\
  \bibinfo {pages} {011} (\bibinfo {year} {2011})},\ \Eprint
  {http://arxiv.org/abs/1008.1495} {arXiv:1008.1495 [astro-ph.CO]} \BibitemShut
  {NoStop}%
\bibitem [{\citenamefont {Mertens}\ \emph {et~al.}(2015)\citenamefont {Mertens}
  \emph {et~al.}}]{Mertens:2014nha}%
  \BibitemOpen
  \bibfield  {author} {\bibinfo {author} {\bibfnamefont {S.}~\bibnamefont
  {Mertens}} \emph {et~al.},\ }\href {\doibase 10.1088/1475-7516/2015/02/020}
  {\bibfield  {journal} {\bibinfo  {journal} {JCAP}\ }\textbf {\bibinfo
  {volume} {02}},\ \bibinfo {pages} {020} (\bibinfo {year} {2015})},\ \Eprint
  {http://arxiv.org/abs/1409.0920} {arXiv:1409.0920 [physics.ins-det]}
  \BibitemShut {NoStop}%
\bibitem [{\citenamefont {Abada}\ \emph {et~al.}(2019)\citenamefont {Abada},
  \citenamefont {Hern\'andez-Cabezudo},\ and\ \citenamefont
  {Marcano}}]{Abada:2018qok}%
  \BibitemOpen
  \bibfield  {author} {\bibinfo {author} {\bibfnamefont {A.}~\bibnamefont
  {Abada}}, \bibinfo {author} {\bibfnamefont {A.}~\bibnamefont
  {Hern\'andez-Cabezudo}}, \ and\ \bibinfo {author} {\bibfnamefont
  {X.}~\bibnamefont {Marcano}},\ }\href {\doibase 10.1007/JHEP01(2019)041}
  {\bibfield  {journal} {\bibinfo  {journal} {JHEP}\ }\textbf {\bibinfo
  {volume} {01}},\ \bibinfo {pages} {041} (\bibinfo {year} {2019})},\ \Eprint
  {http://arxiv.org/abs/1807.01331} {arXiv:1807.01331 [hep-ph]} \BibitemShut
  {NoStop}%
\bibitem [{\citenamefont {Aker}\ \emph {et~al.}(2020)\citenamefont {Aker} \emph
  {et~al.}}]{Aker:2020vrf}%
  \BibitemOpen
  \bibfield  {author} {\bibinfo {author} {\bibfnamefont {M.}~\bibnamefont
  {Aker}} \emph {et~al.} (\bibinfo {collaboration} {KATRIN}),\ }\href@noop {}
  {\  (\bibinfo {year} {2020})},\ \Eprint {http://arxiv.org/abs/2011.05087}
  {arXiv:2011.05087 [hep-ex]} \BibitemShut {NoStop}%
\bibitem [{\citenamefont {Holzschuh}\ \emph {et~al.}(2000)\citenamefont
  {Holzschuh}, \citenamefont {Palermo}, \citenamefont {Stussi},\ and\
  \citenamefont {Wenk}}]{Holzschuh:2000nj}%
  \BibitemOpen
  \bibfield  {author} {\bibinfo {author} {\bibfnamefont {E.}~\bibnamefont
  {Holzschuh}}, \bibinfo {author} {\bibfnamefont {L.}~\bibnamefont {Palermo}},
  \bibinfo {author} {\bibfnamefont {H.}~\bibnamefont {Stussi}}, \ and\ \bibinfo
  {author} {\bibfnamefont {P.}~\bibnamefont {Wenk}},\ }\href {\doibase
  10.1016/S0370-2693(00)00476-7} {\bibfield  {journal} {\bibinfo  {journal}
  {Phys. Lett. B}\ }\textbf {\bibinfo {volume} {482}},\ \bibinfo {pages} {1}
  (\bibinfo {year} {2000})}\BibitemShut {NoStop}%
\bibitem [{\citenamefont {Derbin}\ \emph {et~al.}(2018)\citenamefont {Derbin}
  \emph {et~al.}}]{Derbin:2018dbu}%
  \BibitemOpen
  \bibfield  {author} {\bibinfo {author} {\bibfnamefont {A.}~\bibnamefont
  {Derbin}} \emph {et~al.},\ }\href {\doibase 10.1134/S0021364018200067}
  {\bibfield  {journal} {\bibinfo  {journal} {JETP Lett.}\ }\textbf {\bibinfo
  {volume} {108}},\ \bibinfo {pages} {499} (\bibinfo {year}
  {2018})}\BibitemShut {NoStop}%
\bibitem [{\citenamefont {Doi}\ \emph {et~al.}(1981)\citenamefont {Doi} \emph
  {et~al.}}]{Doi:1981mi}%
  \BibitemOpen
  \bibfield  {author} {\bibinfo {author} {\bibfnamefont {M.}~\bibnamefont
  {Doi}} \emph {et~al.},\ }\href {\doibase 10.1143/PTP.66.1739} {\bibfield
  {journal} {\bibinfo  {journal} {Prog. Theor. Phys.}\ }\textbf {\bibinfo
  {volume} {66}},\ \bibinfo {pages} {1739} (\bibinfo {year} {1981})},\ \bibinfo
  {note} {[Erratum: Prog.Theor.Phys. 68, 347 (1982)]}\BibitemShut {NoStop}%
\bibitem [{\citenamefont {Doi}\ and\ \citenamefont
  {others.}(1981)}]{Doi:1981mj}%
  \BibitemOpen
  \bibfield  {author} {\bibinfo {author} {\bibfnamefont {M.}~\bibnamefont
  {Doi}}\ and\ \bibinfo {author} {\bibnamefont {others.}},\ }\href {\doibase
  10.1143/PTP.66.1765} {\bibfield  {journal} {\bibinfo  {journal} {Prog. Theor.
  Phys.}\ }\textbf {\bibinfo {volume} {66}},\ \bibinfo {pages} {1765} (\bibinfo
  {year} {1981})},\ \bibinfo {note} {[Erratum: Prog.Theor.Phys. 68, 348
  (1982)]}\BibitemShut {NoStop}%
\bibitem [{\citenamefont {Doi}\ \emph {et~al.}(1985)\citenamefont {Doi},
  \citenamefont {Kotani},\ and\ \citenamefont {Takasugi}}]{Doi:1985dx}%
  \BibitemOpen
  \bibfield  {author} {\bibinfo {author} {\bibfnamefont {M.}~\bibnamefont
  {Doi}}, \bibinfo {author} {\bibfnamefont {T.}~\bibnamefont {Kotani}}, \ and\
  \bibinfo {author} {\bibfnamefont {E.}~\bibnamefont {Takasugi}},\ }\href
  {\doibase 10.1143/PTPS.83.1} {\bibfield  {journal} {\bibinfo  {journal}
  {Prog. Theor. Phys. Suppl.}\ }\textbf {\bibinfo {volume} {83}},\ \bibinfo
  {pages} {1} (\bibinfo {year} {1985})}\BibitemShut {NoStop}%
\bibitem [{\citenamefont {Haxton}\ and\ \citenamefont
  {Stephenson}(1984)}]{Haxton:1985am}%
  \BibitemOpen
  \bibfield  {author} {\bibinfo {author} {\bibfnamefont {W.}~\bibnamefont
  {Haxton}}\ and\ \bibinfo {author} {\bibfnamefont {G.}~\bibnamefont
  {Stephenson}},\ }\href {\doibase 10.1016/0146-6410(84)90006-1} {\bibfield
  {journal} {\bibinfo  {journal} {Prog. Part. Nucl. Phys.}\ }\textbf {\bibinfo
  {volume} {12}},\ \bibinfo {pages} {409} (\bibinfo {year} {1984})}\BibitemShut
  {NoStop}%
\bibitem [{\citenamefont {Tomoda}(1991)}]{Tomoda:1990rs}%
  \BibitemOpen
  \bibfield  {author} {\bibinfo {author} {\bibfnamefont {T.}~\bibnamefont
  {Tomoda}},\ }\href {\doibase 10.1088/0034-4885/54/1/002} {\bibfield
  {journal} {\bibinfo  {journal} {Rept. Prog. Phys.}\ }\textbf {\bibinfo
  {volume} {54}},\ \bibinfo {pages} {53} (\bibinfo {year} {1991})}\BibitemShut
  {NoStop}%
\bibitem [{\citenamefont {Kotila}\ and\ \citenamefont
  {Iachello}(2013)}]{Kotila:2013gea}%
  \BibitemOpen
  \bibfield  {author} {\bibinfo {author} {\bibfnamefont {J.}~\bibnamefont
  {Kotila}}\ and\ \bibinfo {author} {\bibfnamefont {F.}~\bibnamefont
  {Iachello}},\ }\href {\doibase 10.1103/PhysRevC.87.024313} {\bibfield
  {journal} {\bibinfo  {journal} {Phys. Rev. C}\ }\textbf {\bibinfo {volume}
  {87}},\ \bibinfo {pages} {024313} (\bibinfo {year} {2013})},\ \Eprint
  {http://arxiv.org/abs/1303.4124} {arXiv:1303.4124 [nucl-th]} \BibitemShut
  {NoStop}%
\bibitem [{\citenamefont {Stefanik}\ \emph {et~al.}(2015)\citenamefont
  {Stefanik}, \citenamefont {Simkovic},\ and\ \citenamefont
  {Faessler}}]{Stefanik:2015bya}%
  \BibitemOpen
  \bibfield  {author} {\bibinfo {author} {\bibfnamefont {D.}~\bibnamefont
  {Stefanik}}, \bibinfo {author} {\bibfnamefont {F.}~\bibnamefont {Simkovic}},
  \ and\ \bibinfo {author} {\bibfnamefont {A.}~\bibnamefont {Faessler}},\
  }\href {\doibase 10.1103/PhysRevC.91.064311} {\bibfield  {journal} {\bibinfo
  {journal} {Phys. Rev. C}\ }\textbf {\bibinfo {volume} {91}},\ \bibinfo
  {pages} {064311} (\bibinfo {year} {2015})},\ \Eprint
  {http://arxiv.org/abs/1506.00835} {arXiv:1506.00835 [nucl-th]} \BibitemShut
  {NoStop}%
\bibitem [{\citenamefont {\v{S}imkovic}\ \emph
  {et~al.}(2018{\natexlab{a}})\citenamefont {\v{S}imkovic}, \citenamefont
  {Dvornick\'y}, \citenamefont {Stef\'anik},\ and\ \citenamefont
  {Faessler}}]{Simkovic:2018rdz}%
  \BibitemOpen
  \bibfield  {author} {\bibinfo {author} {\bibfnamefont {F.}~\bibnamefont
  {\v{S}imkovic}}, \bibinfo {author} {\bibfnamefont {R.}~\bibnamefont
  {Dvornick\'y}}, \bibinfo {author} {\bibfnamefont {D.}~\bibnamefont
  {Stef\'anik}}, \ and\ \bibinfo {author} {\bibfnamefont {A.}~\bibnamefont
  {Faessler}},\ }\href {\doibase 10.1103/PhysRevC.97.034315} {\bibfield
  {journal} {\bibinfo  {journal} {Phys. Rev. C}\ }\textbf {\bibinfo {volume}
  {97}},\ \bibinfo {pages} {034315} (\bibinfo {year} {2018}{\natexlab{a}})},\
  \Eprint {http://arxiv.org/abs/1804.04227} {arXiv:1804.04227 [nucl-th]}
  \BibitemShut {NoStop}%
\bibitem [{\citenamefont {Primakoff}\ and\ \citenamefont
  {Rosen}(1959)}]{Primakoff:1959chj}%
  \BibitemOpen
  \bibfield  {author} {\bibinfo {author} {\bibfnamefont {H.}~\bibnamefont
  {Primakoff}}\ and\ \bibinfo {author} {\bibfnamefont {S.}~\bibnamefont
  {Rosen}},\ }\href {\doibase 10.1088/0034-4885/22/1/305} {\bibfield  {journal}
  {\bibinfo  {journal} {Rept. Prog. Phys.}\ }\textbf {\bibinfo {volume} {22}},\
  \bibinfo {pages} {121} (\bibinfo {year} {1959})}\BibitemShut {NoStop}%
\bibitem [{\citenamefont {Abgrall}\ \emph {et~al.}(2017)\citenamefont {Abgrall}
  \emph {et~al.}}]{Abgrall:2017syy}%
  \BibitemOpen
  \bibfield  {author} {\bibinfo {author} {\bibfnamefont {N.}~\bibnamefont
  {Abgrall}} \emph {et~al.} (\bibinfo {collaboration} {LEGEND}),\ }\href
  {\doibase 10.1063/1.5007652} {\bibfield  {journal} {\bibinfo  {journal} {AIP
  Conf. Proc.}\ }\textbf {\bibinfo {volume} {1894}},\ \bibinfo {pages} {020027}
  (\bibinfo {year} {2017})},\ \Eprint {http://arxiv.org/abs/1709.01980}
  {arXiv:1709.01980 [physics.ins-det]} \BibitemShut {NoStop}%
\bibitem [{\citenamefont {Kharusi}\ \emph {et~al.}(2018)\citenamefont {Kharusi}
  \emph {et~al.}}]{Kharusi:2018eqi}%
  \BibitemOpen
  \bibfield  {author} {\bibinfo {author} {\bibfnamefont {S.~A.}\ \bibnamefont
  {Kharusi}} \emph {et~al.} (\bibinfo {collaboration} {nEXO}),\ }\href@noop {}
  {\  (\bibinfo {year} {2018})},\ \Eprint {http://arxiv.org/abs/1805.11142}
  {arXiv:1805.11142 [physics.ins-det]} \BibitemShut {NoStop}%
\bibitem [{\citenamefont {Armstrong}\ \emph {et~al.}(2019)\citenamefont
  {Armstrong} \emph {et~al.}}]{CUPIDInterestGroup:2019inu}%
  \BibitemOpen
  \bibfield  {author} {\bibinfo {author} {\bibfnamefont {W.}~\bibnamefont
  {Armstrong}} \emph {et~al.} (\bibinfo {collaboration} {CUPID}),\ }\href@noop
  {} {\  (\bibinfo {year} {2019})},\ \Eprint {http://arxiv.org/abs/1907.09376}
  {arXiv:1907.09376 [physics.ins-det]} \BibitemShut {NoStop}%
\bibitem [{\citenamefont {Agostini}\ \emph {et~al.}(2015)\citenamefont
  {Agostini} \emph {et~al.}}]{Agostini:2015nwa}%
  \BibitemOpen
  \bibfield  {author} {\bibinfo {author} {\bibfnamefont {M.}~\bibnamefont
  {Agostini}} \emph {et~al.},\ }\href {\doibase 10.1140/epjc/s10052-015-3627-y}
  {\bibfield  {journal} {\bibinfo  {journal} {Eur. Phys. J. C}\ }\textbf
  {\bibinfo {volume} {75}},\ \bibinfo {pages} {416} (\bibinfo {year} {2015})},\
  \Eprint {http://arxiv.org/abs/1501.02345} {arXiv:1501.02345 [nucl-ex]}
  \BibitemShut {NoStop}%
\bibitem [{\citenamefont {Albert}\ \emph {et~al.}(2014)\citenamefont {Albert}
  \emph {et~al.}}]{Albert:2013gpz}%
  \BibitemOpen
  \bibfield  {author} {\bibinfo {author} {\bibfnamefont {J.}~\bibnamefont
  {Albert}} \emph {et~al.} (\bibinfo {collaboration} {EXO-200}),\ }\href
  {\doibase 10.1103/PhysRevC.89.015502} {\bibfield  {journal} {\bibinfo
  {journal} {Phys. Rev. C}\ }\textbf {\bibinfo {volume} {89}},\ \bibinfo
  {pages} {015502} (\bibinfo {year} {2014})},\ \Eprint
  {http://arxiv.org/abs/1306.6106} {arXiv:1306.6106 [nucl-ex]} \BibitemShut
  {NoStop}%
\bibitem [{\citenamefont {Armengaud}\ \emph {et~al.}(2020)\citenamefont
  {Armengaud} \emph {et~al.}}]{Armengaud:2019rll}%
  \BibitemOpen
  \bibfield  {author} {\bibinfo {author} {\bibfnamefont {E.}~\bibnamefont
  {Armengaud}} \emph {et~al.},\ }\href {\doibase
  10.1140/epjc/s10052-020-8203-4} {\bibfield  {journal} {\bibinfo  {journal}
  {Eur. Phys. J. C}\ }\textbf {\bibinfo {volume} {80}},\ \bibinfo {pages} {674}
  (\bibinfo {year} {2020})},\ \Eprint {http://arxiv.org/abs/1912.07272}
  {arXiv:1912.07272 [nucl-ex]} \BibitemShut {NoStop}%
\bibitem [{\citenamefont {Agostini}\ \emph
  {et~al.}(2020{\natexlab{a}})\citenamefont {Agostini} \emph
  {et~al.}}]{Abramov:2019hhh}%
  \BibitemOpen
  \bibfield  {author} {\bibinfo {author} {\bibfnamefont {M.}~\bibnamefont
  {Agostini}} \emph {et~al.} (\bibinfo {collaboration} {GERDA}),\ }\href
  {\doibase 10.1007/JHEP03(2020)139} {\bibfield  {journal} {\bibinfo  {journal}
  {JHEP}\ }\textbf {\bibinfo {volume} {03}},\ \bibinfo {pages} {139} (\bibinfo
  {year} {2020}{\natexlab{a}})},\ \Eprint {http://arxiv.org/abs/1909.02522}
  {arXiv:1909.02522 [nucl-ex]} \BibitemShut {NoStop}%
\bibitem [{\citenamefont {Agostini}\ \emph
  {et~al.}(2020{\natexlab{b}})\citenamefont {Agostini} \emph
  {et~al.}}]{Agostini:2020xta}%
  \BibitemOpen
  \bibfield  {author} {\bibinfo {author} {\bibfnamefont {M.}~\bibnamefont
  {Agostini}} \emph {et~al.} (\bibinfo {collaboration} {GERDA}),\ }\href@noop
  {} {\  (\bibinfo {year} {2020}{\natexlab{b}})},\ \Eprint
  {http://arxiv.org/abs/2009.06079} {arXiv:2009.06079 [nucl-ex]} \BibitemShut
  {NoStop}%
\bibitem [{\citenamefont {Anton}\ \emph {et~al.}(2019)\citenamefont {Anton}
  \emph {et~al.}}]{Anton:2019wmi}%
  \BibitemOpen
  \bibfield  {author} {\bibinfo {author} {\bibfnamefont {G.}~\bibnamefont
  {Anton}} \emph {et~al.} (\bibinfo {collaboration} {EXO-200}),\ }\href
  {\doibase 10.1103/PhysRevLett.123.161802} {\bibfield  {journal} {\bibinfo
  {journal} {Phys. Rev. Lett.}\ }\textbf {\bibinfo {volume} {123}},\ \bibinfo
  {pages} {161802} (\bibinfo {year} {2019})},\ \Eprint
  {http://arxiv.org/abs/1906.02723} {arXiv:1906.02723 [hep-ex]} \BibitemShut
  {NoStop}%
\bibitem [{\citenamefont {Cowan}\ \emph {et~al.}(2011)\citenamefont {Cowan},
  \citenamefont {Cranmer}, \citenamefont {Gross},\ and\ \citenamefont
  {Vitells}}]{Cowan:2010js}%
  \BibitemOpen
  \bibfield  {author} {\bibinfo {author} {\bibfnamefont {G.}~\bibnamefont
  {Cowan}}, \bibinfo {author} {\bibfnamefont {K.}~\bibnamefont {Cranmer}},
  \bibinfo {author} {\bibfnamefont {E.}~\bibnamefont {Gross}}, \ and\ \bibinfo
  {author} {\bibfnamefont {O.}~\bibnamefont {Vitells}},\ }\href {\doibase
  10.1140/epjc/s10052-011-1554-0} {\bibfield  {journal} {\bibinfo  {journal}
  {Eur. Phys. J. C}\ }\textbf {\bibinfo {volume} {71}},\ \bibinfo {pages}
  {1554} (\bibinfo {year} {2011})},\ \bibinfo {note} {[Erratum: Eur.Phys.J.C
  73, 2501 (2013)]},\ \Eprint {http://arxiv.org/abs/1007.1727} {arXiv:1007.1727
  [physics.data-an]} \BibitemShut {NoStop}%
\bibitem [{\citenamefont {Cousins}\ and\ \citenamefont
  {Highland}(1992)}]{Cousins:1991qz}%
  \BibitemOpen
  \bibfield  {author} {\bibinfo {author} {\bibfnamefont {R.~D.}\ \bibnamefont
  {Cousins}}\ and\ \bibinfo {author} {\bibfnamefont {V.~L.}\ \bibnamefont
  {Highland}},\ }\href {\doibase 10.1016/0168-9002(92)90794-5} {\bibfield
  {journal} {\bibinfo  {journal} {Nucl. Instrum. Meth. A}\ }\textbf {\bibinfo
  {volume} {320}},\ \bibinfo {pages} {331} (\bibinfo {year}
  {1992})}\BibitemShut {NoStop}%
\bibitem [{\citenamefont {Bellini}\ \emph {et~al.}(2013)\citenamefont {Bellini}
  \emph {et~al.}}]{Bellini:2013uui}%
  \BibitemOpen
  \bibfield  {author} {\bibinfo {author} {\bibfnamefont {G.}~\bibnamefont
  {Bellini}} \emph {et~al.} (\bibinfo {collaboration} {Borexino}),\ }\href
  {\doibase 10.1103/PhysRevD.88.072010} {\bibfield  {journal} {\bibinfo
  {journal} {Phys. Rev. D}\ }\textbf {\bibinfo {volume} {88}},\ \bibinfo
  {pages} {072010} (\bibinfo {year} {2013})},\ \Eprint
  {http://arxiv.org/abs/1311.5347} {arXiv:1311.5347 [hep-ex]} \BibitemShut
  {NoStop}%
\bibitem [{\citenamefont {Barea}\ \emph {et~al.}(2015)\citenamefont {Barea},
  \citenamefont {Kotila},\ and\ \citenamefont {Iachello}}]{Barea:2015kwa}%
  \BibitemOpen
  \bibfield  {author} {\bibinfo {author} {\bibfnamefont {J.}~\bibnamefont
  {Barea}}, \bibinfo {author} {\bibfnamefont {J.}~\bibnamefont {Kotila}}, \
  and\ \bibinfo {author} {\bibfnamefont {F.}~\bibnamefont {Iachello}},\ }\href
  {\doibase 10.1103/PhysRevC.91.034304} {\bibfield  {journal} {\bibinfo
  {journal} {Phys. Rev. C}\ }\textbf {\bibinfo {volume} {91}},\ \bibinfo
  {pages} {034304} (\bibinfo {year} {2015})},\ \Eprint
  {http://arxiv.org/abs/1506.08530} {arXiv:1506.08530 [nucl-th]} \BibitemShut
  {NoStop}%
\bibitem [{\citenamefont {Men\'endez}(2018)}]{Menendez:2017fdf}%
  \BibitemOpen
  \bibfield  {author} {\bibinfo {author} {\bibfnamefont {J.}~\bibnamefont
  {Men\'endez}},\ }\href {\doibase 10.1088/1361-6471/aa9bd4} {\bibfield
  {journal} {\bibinfo  {journal} {J. Phys. G}\ }\textbf {\bibinfo {volume}
  {45}},\ \bibinfo {pages} {014003} (\bibinfo {year} {2018})},\ \Eprint
  {http://arxiv.org/abs/1804.02105} {arXiv:1804.02105 [nucl-th]} \BibitemShut
  {NoStop}%
\bibitem [{\citenamefont {Horoi}\ and\ \citenamefont
  {Neacsu}(2016)}]{Horoi:2015tkc}%
  \BibitemOpen
  \bibfield  {author} {\bibinfo {author} {\bibfnamefont {M.}~\bibnamefont
  {Horoi}}\ and\ \bibinfo {author} {\bibfnamefont {A.}~\bibnamefont {Neacsu}},\
  }\href {\doibase 10.1103/PhysRevC.93.024308} {\bibfield  {journal} {\bibinfo
  {journal} {Phys. Rev. C}\ }\textbf {\bibinfo {volume} {93}},\ \bibinfo
  {pages} {024308} (\bibinfo {year} {2016})},\ \Eprint
  {http://arxiv.org/abs/1511.03711} {arXiv:1511.03711 [nucl-th]} \BibitemShut
  {NoStop}%
\bibitem [{\citenamefont {L\'opez~Vaquero}\ \emph {et~al.}(2013)\citenamefont
  {L\'opez~Vaquero}, \citenamefont {Rodr\'\i{}guez},\ and\ \citenamefont
  {Egido}}]{Vaquero:2014dna}%
  \BibitemOpen
  \bibfield  {author} {\bibinfo {author} {\bibfnamefont {N.}~\bibnamefont
  {L\'opez~Vaquero}}, \bibinfo {author} {\bibfnamefont {T.~R.}\ \bibnamefont
  {Rodr\'\i{}guez}}, \ and\ \bibinfo {author} {\bibfnamefont {J.~L.}\
  \bibnamefont {Egido}},\ }\href {\doibase 10.1103/PhysRevLett.111.142501}
  {\bibfield  {journal} {\bibinfo  {journal} {Phys. Rev. Lett.}\ }\textbf
  {\bibinfo {volume} {111}},\ \bibinfo {pages} {142501} (\bibinfo {year}
  {2013})},\ \Eprint {http://arxiv.org/abs/1401.0650} {arXiv:1401.0650
  [nucl-th]} \BibitemShut {NoStop}%
\bibitem [{\citenamefont {Rodriguez}\ and\ \citenamefont
  {Martinez-Pinedo}(2010)}]{Rodriguez:2010mn}%
  \BibitemOpen
  \bibfield  {author} {\bibinfo {author} {\bibfnamefont {T.~R.}\ \bibnamefont
  {Rodriguez}}\ and\ \bibinfo {author} {\bibfnamefont {G.}~\bibnamefont
  {Martinez-Pinedo}},\ }\href {\doibase 10.1103/PhysRevLett.105.252503}
  {\bibfield  {journal} {\bibinfo  {journal} {Phys. Rev. Lett.}\ }\textbf
  {\bibinfo {volume} {105}},\ \bibinfo {pages} {252503} (\bibinfo {year}
  {2010})},\ \Eprint {http://arxiv.org/abs/1008.5260} {arXiv:1008.5260
  [nucl-th]} \BibitemShut {NoStop}%
\bibitem [{\citenamefont {Song}\ \emph {et~al.}(2017)\citenamefont {Song},
  \citenamefont {Yao}, \citenamefont {Ring},\ and\ \citenamefont
  {Meng}}]{Song:2017ktj}%
  \BibitemOpen
  \bibfield  {author} {\bibinfo {author} {\bibfnamefont {L.}~\bibnamefont
  {Song}}, \bibinfo {author} {\bibfnamefont {J.}~\bibnamefont {Yao}}, \bibinfo
  {author} {\bibfnamefont {P.}~\bibnamefont {Ring}}, \ and\ \bibinfo {author}
  {\bibfnamefont {J.}~\bibnamefont {Meng}},\ }\href {\doibase
  10.1103/PhysRevC.95.024305} {\bibfield  {journal} {\bibinfo  {journal} {Phys.
  Rev. C}\ }\textbf {\bibinfo {volume} {95}},\ \bibinfo {pages} {024305}
  (\bibinfo {year} {2017})},\ \Eprint {http://arxiv.org/abs/1702.02448}
  {arXiv:1702.02448 [nucl-th]} \BibitemShut {NoStop}%
\bibitem [{\citenamefont {Hyv\"arinen}\ and\ \citenamefont
  {Suhonen}(2015)}]{Hyvarinen:2015bda}%
  \BibitemOpen
  \bibfield  {author} {\bibinfo {author} {\bibfnamefont {J.}~\bibnamefont
  {Hyv\"arinen}}\ and\ \bibinfo {author} {\bibfnamefont {J.}~\bibnamefont
  {Suhonen}},\ }\href {\doibase 10.1103/PhysRevC.91.024613} {\bibfield
  {journal} {\bibinfo  {journal} {Phys. Rev. C}\ }\textbf {\bibinfo {volume}
  {91}},\ \bibinfo {pages} {024613} (\bibinfo {year} {2015})}\BibitemShut
  {NoStop}%
\bibitem [{\citenamefont {Mustonen}\ and\ \citenamefont
  {Engel}(2013)}]{Mustonen:2013zu}%
  \BibitemOpen
  \bibfield  {author} {\bibinfo {author} {\bibfnamefont {M.}~\bibnamefont
  {Mustonen}}\ and\ \bibinfo {author} {\bibfnamefont {J.}~\bibnamefont
  {Engel}},\ }\href {\doibase 10.1103/PhysRevC.87.064302} {\bibfield  {journal}
  {\bibinfo  {journal} {Phys. Rev. C}\ }\textbf {\bibinfo {volume} {87}},\
  \bibinfo {pages} {064302} (\bibinfo {year} {2013})},\ \Eprint
  {http://arxiv.org/abs/1301.6997} {arXiv:1301.6997 [nucl-th]} \BibitemShut
  {NoStop}%
\bibitem [{\citenamefont {Fang}\ \emph {et~al.}(2018)\citenamefont {Fang},
  \citenamefont {Faessler},\ and\ \citenamefont {Simkovic}}]{Fang:2018tui}%
  \BibitemOpen
  \bibfield  {author} {\bibinfo {author} {\bibfnamefont {D.-L.}\ \bibnamefont
  {Fang}}, \bibinfo {author} {\bibfnamefont {A.}~\bibnamefont {Faessler}}, \
  and\ \bibinfo {author} {\bibfnamefont {F.}~\bibnamefont {Simkovic}},\ }\href
  {\doibase 10.1103/PhysRevC.97.045503} {\bibfield  {journal} {\bibinfo
  {journal} {Phys. Rev. C}\ }\textbf {\bibinfo {volume} {97}},\ \bibinfo
  {pages} {045503} (\bibinfo {year} {2018})},\ \Eprint
  {http://arxiv.org/abs/1803.09195} {arXiv:1803.09195 [nucl-th]} \BibitemShut
  {NoStop}%
\bibitem [{\citenamefont {\v{S}imkovic}\ \emph
  {et~al.}(2018{\natexlab{b}})\citenamefont {\v{S}imkovic}, \citenamefont
  {Smetana},\ and\ \citenamefont {Vogel}}]{Simkovic:2018hiq}%
  \BibitemOpen
  \bibfield  {author} {\bibinfo {author} {\bibfnamefont {F.}~\bibnamefont
  {\v{S}imkovic}}, \bibinfo {author} {\bibfnamefont {A.}~\bibnamefont
  {Smetana}}, \ and\ \bibinfo {author} {\bibfnamefont {P.}~\bibnamefont
  {Vogel}},\ }\href {\doibase 10.1103/PhysRevC.98.064325} {\bibfield  {journal}
  {\bibinfo  {journal} {Phys. Rev. C}\ }\textbf {\bibinfo {volume} {98}},\
  \bibinfo {pages} {064325} (\bibinfo {year} {2018}{\natexlab{b}})},\ \Eprint
  {http://arxiv.org/abs/1808.05016} {arXiv:1808.05016 [nucl-th]} \BibitemShut
  {NoStop}%
\bibitem [{\citenamefont {Coraggio}\ \emph {et~al.}(2020)\citenamefont
  {Coraggio} \emph {et~al.}}]{Coraggio:2020hwx}%
  \BibitemOpen
  \bibfield  {author} {\bibinfo {author} {\bibfnamefont {L.}~\bibnamefont
  {Coraggio}} \emph {et~al.},\ }\href {\doibase 10.1103/PhysRevC.101.044315}
  {\bibfield  {journal} {\bibinfo  {journal} {Phys. Rev. C}\ }\textbf {\bibinfo
  {volume} {101}},\ \bibinfo {pages} {044315} (\bibinfo {year} {2020})},\
  \Eprint {http://arxiv.org/abs/2001.00890} {arXiv:2001.00890 [nucl-th]}
  \BibitemShut {NoStop}%
\bibitem [{\citenamefont {Vincent}\ \emph {et~al.}(2015)\citenamefont
  {Vincent}, \citenamefont {Martinez}, \citenamefont {Hern\'andez},
  \citenamefont {Lattanzi},\ and\ \citenamefont {Mena}}]{Vincent:2014rja}%
  \BibitemOpen
  \bibfield  {author} {\bibinfo {author} {\bibfnamefont {A.~C.}\ \bibnamefont
  {Vincent}}, \bibinfo {author} {\bibfnamefont {E.~F.}\ \bibnamefont
  {Martinez}}, \bibinfo {author} {\bibfnamefont {P.}~\bibnamefont
  {Hern\'andez}}, \bibinfo {author} {\bibfnamefont {M.}~\bibnamefont
  {Lattanzi}}, \ and\ \bibinfo {author} {\bibfnamefont {O.}~\bibnamefont
  {Mena}},\ }\href {\doibase 10.1088/1475-7516/2015/04/006} {\bibfield
  {journal} {\bibinfo  {journal} {JCAP}\ }\textbf {\bibinfo {volume} {04}},\
  \bibinfo {pages} {006} (\bibinfo {year} {2015})},\ \Eprint
  {http://arxiv.org/abs/1408.1956} {arXiv:1408.1956 [astro-ph.CO]} \BibitemShut
  {NoStop}%
\bibitem [{\citenamefont {Sabti}\ \emph {et~al.}(2020)\citenamefont {Sabti},
  \citenamefont {Alvey}, \citenamefont {Escudero}, \citenamefont {Fairbairn},\
  and\ \citenamefont {Blas}}]{Sabti:2019mhn}%
  \BibitemOpen
  \bibfield  {author} {\bibinfo {author} {\bibfnamefont {N.}~\bibnamefont
  {Sabti}}, \bibinfo {author} {\bibfnamefont {J.}~\bibnamefont {Alvey}},
  \bibinfo {author} {\bibfnamefont {M.}~\bibnamefont {Escudero}}, \bibinfo
  {author} {\bibfnamefont {M.}~\bibnamefont {Fairbairn}}, \ and\ \bibinfo
  {author} {\bibfnamefont {D.}~\bibnamefont {Blas}},\ }\href {\doibase
  10.1088/1475-7516/2020/01/004} {\bibfield  {journal} {\bibinfo  {journal}
  {JCAP}\ }\textbf {\bibinfo {volume} {01}},\ \bibinfo {pages} {004} (\bibinfo
  {year} {2020})},\ \Eprint {http://arxiv.org/abs/1910.01649} {arXiv:1910.01649
  [hep-ph]} \BibitemShut {NoStop}%
\bibitem [{\citenamefont {Bolton}\ \emph
  {et~al.}(2020{\natexlab{b}})\citenamefont {Bolton}, \citenamefont {Deppisch},
  \citenamefont {Gr\'af},\ and\ \citenamefont {\v{S}imkovic}}]{Bolton:2020ncv}%
  \BibitemOpen
  \bibfield  {author} {\bibinfo {author} {\bibfnamefont {P.~D.}\ \bibnamefont
  {Bolton}}, \bibinfo {author} {\bibfnamefont {F.~F.}\ \bibnamefont
  {Deppisch}}, \bibinfo {author} {\bibfnamefont {L.}~\bibnamefont {Gr\'af}}, \
  and\ \bibinfo {author} {\bibfnamefont {F.}~\bibnamefont {\v{S}imkovic}},\
  }\href@noop {} {\  (\bibinfo {year} {2020}{\natexlab{b}})},\ \Eprint
  {http://arxiv.org/abs/2011.13387} {arXiv:2011.13387 [hep-ph]} \BibitemShut
  {NoStop}%
\end{thebibliography}%

\end{document}